\documentclass{ws-ijtaf}

\begin{document}

\markboth{Authors' Names}
{Instructions for Typing Manuscripts (Paper's Title)}

%
\catchline{}{}{}{}{}
%

\title{Pricing of options on stocks that are driven by multi-dimensional 
coupled price-temporal infinitely divisible fluctuations.} 
\author{Przemys\mbox{\l}aw Repetowicz}
\address{Department of Physics, Trinity College Dublin 2, Ireland}
\email{repetowp@tcd.ie}

\author{Mark M. Meerschaert}
\address{Department of Physics 
211 Leifson Physics Building  
University of Nevada         
Reno NV 8955}
\email{mcubed@unr.edu}

\author{Peter Richmond}
\address{Department of Physics, Trinity College Dublin 2, Ireland}
\email{richmond@tcd.ie}
\maketitle


\begin{history}
\received{(29/09/2004)}
\revised{(Day Month Year)}
\end{history}

\begin{abstract}
We model the price of a stock via a Lang\'{e}vin equation
with multi-dimensional fluctuations coupled both in the price
in time. We generalise previous models in that we assume that 
the fluctuations conditioned on the time step
are assumed to be 
compound Poisson processes with operator stable jump intensities.
We derive exact relations for Fourier transforms of the jump intensity
in case of different scaling indices $\underline{\underline{E}}$ 
of the process.
We express the Fourier transform of the joint probability density
of the process to attain given values at several different times
and to attain a given maximal value in a given time period
through Fourier transforms of the jump intensity.
Then we consider a portfolio composed of stocks and of options
on stocks and we derive the Fourier transform 
of a random variable $\mathfrak{D}_t$ (deviation of the portfolio)
that is defined as a small temporal change of the portfolio
diminished by the the compound interest earned.
We derive a functional equation for the
price of the option on the stock as a function of the stock
subject to the condition that
the deviation of the portfolio has a zero mean
$E\left[ \mathfrak{D}_t \right] = 0$ 
Therefore the option pricing problem may have a solution.
\end{abstract}

\section{Introduction}
Early statistical models of financial markets assumed a Gaussian distribution \cite{Bach}.
However, there is evidence \cite{GopiMeyer} that stock price returns
are not Gauss distributed and that the distributions
diminish slowly in the high end (fat tails).
A better statistical description 
is provided by a model where the logarithm of the stock price
is a one dimensional compound Poisson process with jumps being 
asymptotically power-law distributed (stable L\'{e}vy process \cite{Meerschaert}).
That model, however, does not eliminate discrepancies between experimental facts 
and the theory and drawbacks of the theory. 

The exponents of the power laws (tail indices) in distributions appear to be
outside of the range of validity predicted by the theory \cite{GopiMeyer}.
Furthermore models in the literature \cite{Mandel,Fama} assume that jumps in the price are independent
of waiting times between jumps (no price-temporal coupling).

Fat tails can be explained by models basing on operator
stable L\'{e}vy processes \cite{Meerschaert}.
Here the price of the stock is an exponential from a sum of a large number 
of projections of many-dimensional jumps whose distributions are invariant under auto-convolution. 
This implies that the distributions are, in the high end, mixtures
of power laws with different exponents, 
each one describing a walk in one of the dimensions.

We briefly describe development of models that account for a price-temporal
coupling (coupled random walks) below.
\cite{RepRich,Masoliver} formulate an analytical model
that yields probability distributions of price jumps under an ``ad hoc'' ansatz
about the form of price-temporal coupling. 
\cite{MeerschaertCoupledI} 
derives a long time scaling limit for the jump probability distribution
in coupled random walks for an anomalous fractional diffusion.
\cite{MeerschaertCopledPRE} obtains results for master equations 
of coupled random walks in the long time scaling limit and show their connection
to fractional derivative calculus.
\cite{KutnerWeierstrassFlights} 
considers one-dimensional continuous-time coupled random walks 
(Weierstrass flights) and derives the long time scaling limit of the mean 
squared displacement of the walker as a function of two parameters 
$\alpha$ and $\beta$  which are fractional dimensions
that represent the flight and the waiting.
An extension  of Weierstrass flights
to a model where the walker, 
after random waiting times, moves with arbitrary finite velocities to new positions
is formulated in \cite{KutnerLevyWalksVaryVel}
and formulas are derived for the sojourn probability density
and for likelihoods of extreme events.
Finally \cite{Raberto,RepRich} construct contingency tables
between logarithmic price returns and waiting times for stocks
on the New York Stock Exchange and for United States Dollar 
-- Japanese Yen exchange rates and reject the hypothesis
of the independence of returns on waiting times
at a high confidence level. 

We formulate a multi-dimensional random walk model  that recognises 
that the jumps are not independent of the waiting times.
The model has a two-level structure 
with the higher level being a walk with random fluctuations preceded 
by random waiting times 
and coupled to those and the lower level describing the fluctuations
as sums of random numbers of random jumps 
with a jump intensity (L\'{e}vy density) $\varpi(x|t)$.
The levels of the model have a subordinated time horizon that will be 
explained in the next section. 
Our model works with two arbitrary functions, viz the L\'{e}vy density $\varpi(x|t)$ 
with a scaling index $\underline{\underline{E}}$
and a waiting time probability density function (pdf) $\rho_{\Delta T}(t)$ 
with a scaling index $0 < \beta \le 1$ and generalises previous models
proposed in the literature in this subject matter. 

\section{The model}
In this section we formulate an assumption about the price evolution of the 
stock and we recollect some definitions from the theory of stochastic processes.
In section \ref{sec:TheStockPrice}  
we recapitulate some known facts concerned with limit distributions
of sums of independent random vectors \cite{Meerschaert}. 
In sections \ref{sec:Scal} and \ref{sec:ScalRot} we obtain exact relations for Fourier transforms 
of jump intensities and for Fourier transforms of 
distributions of integer powers of one dimensional projections
of jumps that are sums of large numbers of independent random vectors.
In section \ref{eq:JointPdfs} we derive the Fourier transform of 
a joint probability density of an event that the stochastic process 
attains given values at given times and, in addition to that, that it attains
a given maximal value in some predefined time interval.
In section (\ref{sec:OptionPrice}) we define an option on the stock,
define a portfolio that is composed of stocks and of options on stocks,
define a random variable (deviation of portfolio) 
that describes the temporal change of that portfolio in a small time span
diminished by a compound rate of interest earned on the portfolio in that time span.
and we derive conditions on the deviation of the portfolio to have
a zero mean. 

\subsection{The stock price\label{sec:TheStockPrice}}   
Let $\log(S_t)$ be the logarithm of the price of the stock (log-price) at time $t$.
We assume that the log-price has a drift $\alpha$.
In a small time interval $\delta t$ the log-price may or may not change
with respect to this drift.
The probability of that change
depends on a presumed
waiting time pdf $\rho_{\Delta T}(t)$. 
This means that:
\begin{equation}
\log(S_{t+\delta t}) - \log(S_t) = \alpha (\delta t)^\beta +
\left\{
\begin{array}{cc}
 \sum_{i=1}^d \sigma_i \delta L^{(i)}_t 
  &  \mbox{with probability  $\int_0^{\delta t} \rho_{\Delta T}(\xi) d\xi$} \\
0 &  \mbox{with probability  $1 - \int_0^{\delta t} \rho_{\Delta T}(\xi) d\xi$}
\end{array}
\right.
\label{eq:StockPriceFluct}
\end{equation}
where $\beta$ is a real number that will be specified later,
the parameters $\alpha$ and 
$\vec{\sigma} := \left(\sigma_1,\dots,\sigma_d\right)$ (volatility)
are assumed to be constant as a function of time $t$ and of the stock price $S_t$.
We assume that the random vector $\vec{\delta L_t}$ (fluctuation)
is infinitely divisible (see e.g. \cite{Meerschaert} for the precise definition).
This implies that the logarithm of the Fourier transform (log-characteristic function) of it
has a unique 
representation \cite{Meerschaert} in terms of the L\'{e}vy-Khintchin formula.
This means that:
\begin{equation}
\chi_{\vec{\delta L_t}}(\vec{k}) := 
E\left[\exp(\imath \vec{k}\cdot\vec{\delta L_t})\right]
= \imath \vec{a} \cdot \vec{k} - \frac{1}{2} Q(\vec{k},t) +
\int_{\vec{x}\ne 0} \left( e^{\imath \vec{k} \cdot \vec{x}} 
- 1 - \frac{\imath \vec{k} \cdot \vec{x}}{1 + ||\vec{x}||^2} \right) 
\phi(\vec{x}|t) d\vec{x}
\end{equation}
where $\vec{a}$ is a constant, $Q(\vec{k},t)$ is a quadratic form that describes
Gaussian fluctuations
and the integral
represents a compound Poisson process with some rate $c>0$ and 
with a jump pdf $\phi(\vec{x}|t)$.
Here we do not use the measure-theory notation but instead
we replace the measure by Riemann integrals with some probability densities.
We denote $\vec{\delta L_t} = \mathcal{ID}\left[a,Q,\phi\right]$.
Since in our model (\ref{eq:StockPriceFluct}) the drift of the stock price
is already accounted for by the parameter $\alpha$ 
and we neglect the Gaussian part $Q(\vec{k},t)$ in the fluctuations.
We also assume that $\phi(\vec{x}|t)$ is symmetric in $\vec{x}$.
Therefore we have:
\begin{equation}
\vec{\delta L_t} = \mathcal{ID}\left[0,0,\phi\right]
\label{eq:FluctDef}
\end{equation}
The probability density function of fluctuations conditioned on the random waiting time 
$\delta T$  reads:
\begin{equation}
P\left( \vec{\delta L_t} = \delta \vec{l}_t | \delta T = \delta t\right) =
\omega_{\vec{\delta L_t} | \delta T} \left( \delta \vec{l}_t| \delta t \right) 
= \exp(-c) \sum_{n=0}^\infty \frac{c^n}{n!} \phi^{n \otimes}(\delta \vec{l}_t| \delta t) 
\label{eq:FluctConditPdf}
\end{equation}
where $n \otimes$ means an $n$-times auto-convolution. 
The last equality in (\ref{eq:FluctConditPdf}) represents a probability density function
of a compound Poisson process with rate $c$ and jump intensity 
$\phi(\vec{\delta l}_t| \delta t)$.

The joint pdf $\omega_{\vec{\delta L_t},  \delta T}(\delta \vec{l}_t, \delta t)$ 
of a fluctuation $\delta \vec{L}_t = \delta \vec{l}_t$ 
preceded by a waiting $\delta T = \delta t$
reads:
\begin{equation}
\omega_{\vec{\delta L_t},  \delta T}(\delta \vec{l}_t, \delta t) =
\omega_{\vec{\delta L_t} | \delta T} \left( \delta \vec{l}_t| \delta t \right)
\rho_{\delta T}(\delta t)
\end{equation}

Before proceeding further we summarise the definitions 
in Table \ref{tab:SummaryDef}.

\begin{table}
\begin{tabular}{|c|c|}\hline
\begin{minipage}{0.4\textwidth}
\noindent[$\mathcal{A}$] The joint pdf of a fluctuation 
$\delta \vec{L}_t = \delta \vec{l}_t$ 
preceded by a waiting time $\delta T = \delta t$
(the joint fluctuation pdf)
\end{minipage} 
&
\begin{minipage}{0.4\textwidth}
\[
\omega_{\vec{\delta L_t},  \delta T}(\delta \vec{l}_t, \delta t)
\]
\end{minipage}
\\ \hline
\begin{minipage}{0.4\textwidth}
\noindent[$\mathcal{B}$] The pdf of a fluctuation 
$\delta \vec{L}_t = \delta \vec{l}_t$ 
conditioned on a random waiting time $\delta T = \delta t$
(the conditional fluctuation pdf)
\end{minipage} 
&
\begin{minipage}{0.4\textwidth}
\[
\omega_{\vec{\delta L_t}|  \delta T}(\delta \vec{l}_t| \delta t)
\]
\end{minipage}
\\ \hline
\begin{minipage}{0.4\textwidth}
\noindent[$\mathcal{C}$] The joint pdf of a jump 
$\delta \vec{\mathfrak{L}}_t = \delta \vec{\mathfrak{l}}_t$
preceded by a random waiting time $\delta T = \delta t$
(the joint jump pdf)
\end{minipage} 
&
\begin{minipage}{0.4\textwidth}
\[
\phi(\delta \vec{\mathfrak{l}}_t, \delta t)
\]
\end{minipage}
\\ \hline
\begin{minipage}{0.4\textwidth}
\noindent[$\mathcal{D}$] The pdf of a jump 
$\delta \vec{\mathfrak{L}}_t$ conditioned on a 
random waiting time $\delta T = \delta t$
(the conditional jump pdf)
\end{minipage} 
&
\begin{minipage}{0.4\textwidth}
\[
\phi(\delta \vec{\mathfrak{l}}_t| \delta t)
\]
\end{minipage}
\\ \hline
\begin{minipage}{0.4\textwidth}
\noindent[$\mathcal{E}$] The pdf of the $n^{\mbox{th}}$ power
$\Xi^{(n)} := \left(\vec{\sigma} \cdot \delta \vec{L}_t \right)^n = z$
of the fluctuation term in (\ref{eq:StockPriceFluct})
conditioned on the waiting time $\delta T = \delta t$
(the conditional $n$-marginal pdf)
\end{minipage} 
&
\begin{minipage}{0.4\textwidth}
\[
\nu^{(n)}(z | \delta t)
\]
\end{minipage}
\\ \hline
\begin{minipage}{0.4\textwidth}
\noindent[$\mathcal{F}$] The L\'{e}vy measure of jumps
(the jump intensity)
\end{minipage} 
&
\begin{minipage}{0.4\textwidth}
\[
\varpi(\vec{\delta l_t})
\]
\end{minipage}
\\ \hline
\begin{minipage}{0.4\textwidth}
\noindent[$\mathcal{G}$] The pdf of a waiting time $\delta T = \delta t$
(the waiting time pdf)
\end{minipage} 
&
\begin{minipage}{0.4\textwidth}
\[
\rho_{\delta T}(\delta t)
\]
\end{minipage}
\\ \hline
\end{tabular}
\caption{Summary of definitions\label{tab:SummaryDef}}
\end{table}
Note that the quantities [$\mathcal{A}$], [$\mathcal{B}$] 
differ from their counterparts [$\mathcal{C}$], [$\mathcal{D}$] 
in that, since the fluctuations are compound Poisson random variables,
a fluctuation $\delta \vec{L}_t$ consists, in general, of several jumps
$\delta \vec{L}_t = \sum_{j=1}^{N_t} \vec{\delta \mathfrak{L}_t^{(j)}}$
where the number of jumps $N_t$ is a Poisson random number $N_t = \mbox{Poisson(c)}$
with rate $c$.
Therefore the quantities  [$\mathcal{A}$], [$\mathcal{B}$] 
are expressed via sums of auto-convolutions of  [$\mathcal{C}$], [$\mathcal{D}$] 
according to formula (\ref{eq:FluctConditPdf}).
We also stress that all statements in this paper are concerned with 
large $\vec{x}$ scaling limits (the high-end of the distribution) 
of distributions in question.

We assume that the time horizon $\delta T$ and the price fluctuations $\vec{\delta L_t}$
are small at the higher level of the model and large at the lower level.
Note that such an assumption, accompanied by additional ad hoc hypotheses,
was tacit in previous works 
(equation (16) in \cite{Masoliver} cond-mat/0308017 preprint).
This assumption will enable us to use the scaling limit 
of the jump probability density function in a coupled random walk:
This is the Theorem 2.2 on page 733 in \cite{MeerschaertCoupledI} that reads: 
\begin{theorem}
Assume that the waiting time is $\beta$-stable with 
index $0 < \beta < 1$, meaning that the Laplace transform
of the waiting time pdf has a following form:
$\mathcal{L}_t\left[ \rho \right](s)=\exp\left(-\mathcal{K} \Gamma(1 - \beta) s^\beta\right)$
where $\mathcal{K} = \mathcal{K}(a)$ is a normalisation constant of the waiting time pdf
($\mathcal{K}(a) = \int_{a}^\infty \rho(t) dt$) for some $a >> 0 $.

The large $\vec{x}$ and large $t$ scaling limit of the 
of the joint jump pdf in a coupled random walk reads
\begin{equation}
\phi(\vec{x}, t) = \mbox{det$(t^{-\beta \underline{\underline{E}}})$}
                      \varpi (t^{- \beta \underline{\underline{E}}}\vec{x}) 
\frac{\mathcal{K}(a)\beta}{t^{\beta + 1}}
\label{eq:ScalinglimitTh}
\end{equation}
where $\varpi: \mathbb{R}^d \ni \vec{x} \rightarrow \varpi(\vec{x}) \in \mathbb{R}_+$ 
is the probability density in $\vec{x}$ (termed as the jump intensity), 
and the matrix $\underline{\underline{E}}$ is an exponent of the operator stable
law (stable index) corresponding to the jump $\vec{x}\in \mathbb{R}^d$ random variable
and $\mbox{det$(t^{-\beta \underline{\underline{E}}})$}$
is a normalisation constant of the jump pdf.

The eigenvalues $\lambda$ of $\underline{\underline{E}}$ satisfy
$\mbox{Re$[\lambda] \ge 1/2$}$ and the eigenvalues $\mbox{Re$[\lambda] = 1/2$}$ are simple 
(not degenerate).
\end{theorem}

\begin{corollary}
The scaling limit of the conditional jump pdf is:
\begin{equation}
\phi(\vec{x}| t) = 
\frac{\phi(\vec{x}, t)}{\int\limits_{\vec{x} \in \mathbb{R}^d} 
\phi(\vec{x}, t) d\vec{x}}
= \mbox{det$(t^{-\beta \underline{\underline{E}}})$}
                      \varpi (t^{- \beta \underline{\underline{E}}}\vec{x})
\label{eq:ScalingLimitCor}
\end{equation}
This follows in a straightforward manner by integrating over $\vec{x}$
in (\ref{eq:ScalinglimitTh}).
\end{corollary}
\begin{corollary}
The jump intensity $\varpi$ satisfies a functional equation:
\begin{equation}
\exists_{\underline{\underline{E}} \in \mathbb{L}(\mathbb{R}^d)}
\forall \mathop{\vec{x}\ne \vec{0}}_{t \ne 0}
\quad\quad
\varpi^{t \otimes}(\vec{x}) = 
\varpi(t^{-\underline{\underline{E}}}\vec{x}) \mbox{det$(t^{-\underline{\underline{E}}})$}
\label{eq:ISExpDef}
\end{equation}
where $\varpi^{t \otimes}$ is the $t$th auto-convolution 
defined as a limit of $t_n$-times
auto-convolutions for some rational $t_n$  such that 
$\mbox{lim}_{n\rightarrow \infty} t_n = t$. 
Since the random process is infinitely divisible 
the definition is correct.

The corollary follows from the self-similarity of the random walk.
This means that there exists a sequence of linear operators 
$\underline{\underline{B}}(n) \in \mathbb{L}(\mathbb{R}^d)$
and a sequence of real numbers $b_n$ and a sequence 
of integer numbers $k_n$
such that for some independent, identically distributed
random variables $\vec{Y}_i \in \mathbb{R}^d$ we have:
\begin{eqnarray}
\underline{\underline{ B}}(n) \left( \vec{Y}_1 + \dots \vec{Y}_{k_n} \right) - b_{k_n} 
\mathop{\rightarrow}_{n \rightarrow \infty} \vec{Y} 
\quad\mbox{with a pdf $\varpi$} 
\label{eq:CTRWDefI}
\end{eqnarray}
On the other hand if we take the sequence of rational numbers $t_n$ we have
\begin{eqnarray}
\underline{\underline{ B}}(n) \left( \vec{Y}_1 + \dots \vec{Y}_{t_n k_n} \right) 
- b_{t_n k_n} 
\mathop{\rightarrow}_{n \rightarrow \infty} \vec{Y}_1 
\quad\mbox{with a pdf $\varpi^{t \otimes}$} 
\label{eq:CTRWDefII}
\end{eqnarray}
Now if we take $n$ large enough we can write 
$k_n t_n = k_{m(n)}$ for a certain subsequence $m(n)$.
Therefore the random variable $Y_1$ has the same pdf $\varpi$
except that the argument of the pdf is linearly transformed
$ \vec{x} \rightarrow t^{-\underline{\underline{E}}} \vec{x}$.
\end{corollary}

\begin{corollary}
The Fourier transform $\tilde{\varpi}(\vec{k}) := \mathcal{F}_{\vec{x}}[\varpi](\vec{k})$
of the jump intensity $\varpi$ satisfies a functional equation:
\begin{equation}
\tilde{\varpi}^t(\vec{k}) = \tilde{\varpi}( t^{\underline{\underline{E}}^{T}} \vec{k} ) 
\label{eq:ScalingLimitCorFourTransf}
\end{equation}
Where $\underline{\underline{E}}^{T}$ is the transpose of 
$\underline{\underline{E}}$. The right hand side of the equality 
(\ref{eq:ScalingLimitCorFourTransf})
follows in a straightforward manner by taking a Fourier transform of 
the right hand side of (\ref{eq:ISExpDef}).
The left hand side is also straightforward.
\begin{equation}
\mathcal{F}_{\vec{x}} \left[ \varpi^{t \otimes} \right](\vec{k}) =
\mbox{lim}_{n\rightarrow \infty} 
\mathcal{F}_{\vec{x}} \left[ \varpi^{t_n \otimes} \right](\vec{k}) =
\mbox{lim}_{n\rightarrow \infty} 
\tilde{\varpi}^{t_n}(\vec{k}) = \tilde{\varpi}^t(\vec{k})
\end{equation}
see \cite{Meerschaert} for details.
\end{corollary}

We illustrate {\bf Theorem 2.1} in following examples:

\noindent{\bf Example 1.} Let $d=1$ and $\vec{Y}_t = Y_t = \mbox{Normal$(0,t)$}$
is normal with mean zero and variance $t$. Assume that $\beta = $.
Then $\underline{\underline{E}} = E = 1/2$ and 
$\varpi(x) = (2 \pi)^{-1/2} \exp(-x^2/2)$.

\noindent{\bf Example 2.} Let $d=1$ and let the log-characteristic function
of $Y_t$ be $\log(E[\exp(\imath k Y)] = -t|k|^\mu$ where $0 \le \mu \le 2$. 
Then $E = 1/\mu$ and the function $\varpi$ can has an asymptotic expansion:
\begin{equation}
\varpi(x) = \frac{1}{\pi} 
\sum_{n=1}^\infty \sin(\frac{\pi}{2} \mu n) \frac{\Gamma(\mu n + 1)}{n!} \frac{1}{x^{\mu n + 1}}
\label{eq:AsymptExp}
\end{equation}

In the next section we are going to analyse the Langevin
equation (\ref{eq:StockPriceFluct}) for the log-price of the stock driven
by operator stable fluctuations. For this purpose we need the time
dependence of the conditional jump pdf on the waiting time for small waiting times
and we need the probability distribution of the $n^{\mbox{th}}$ power 
of a scalar product $\Xi^{(n)}:= \left(\vec{\sigma} \cdot \delta \vec{L}_t \right)^n$
for $n \in \mathbb{N}$ (the conditional $n$-marginal pdf).
We have:
\begin{proposition}
The conditional jump pdf $\phi(\vec{x}|t)$ depends on time $t$ as follows:
\begin{equation}
\phi(\vec{x}|t) = t^{-\beta \mbox{Tr$ \underline{\underline{ E}}$}}
\varpi (t^{- \beta \underline{\underline{E}}}\vec{x})
\label{eq:TimeDependCondJumpPdf}
\end{equation}
This is a straightforward conclusion from (\ref{eq:ScalingLimitCor}) and from the fact that
$\mbox{det$\left[ e^{\underline{\underline{ A}}} \right] 
= e^{\mbox{Tr$\left[ \underline{\underline{ A}} \right]$}}$}$
for every linear operator $\underline{\underline{ A}} \in \mathbb{L}(\mathbb{R}^d)$.  
\end{proposition}
\begin{proposition}
The Fourier transform of the conditional $n$-marginal probability density function
$\tilde{\nu}^{(n)}(k | \delta t) 
:= \mathcal{F}_{z}\left[\nu^{(n)}\right](k)$
as a function of the Fourier transform of the jump intensity 
$\tilde{\varpi}(\vec{\lambda})
:= \mathcal{F}_{\vec{x}}\left[\varpi\right](\vec{\lambda})$ 
takes following form: 
\begin{eqnarray}
\tilde{\nu}^{(n)}(k | \delta t) = 
\int\limits_{-\infty}^\infty d\mathfrak{l}
\tilde{\varpi}((\delta t)^{\beta \underline{\underline{E}}^{T}} 
\frac{\vec{\sigma}}{|\vec{\sigma}|}(\mathfrak{l}))
\mathcal{K}^{(n)}(\mathfrak{k}, \mathfrak{l})
\label{eq:FourTrNMarginalProp}
\end{eqnarray}
where
\begin{eqnarray}
 \mathcal{K}^{(n)}(\mathfrak{k}, \mathfrak{l}) :=
\left\{
\begin{array}{rr}
2 Re\left[ 
e^{\imath \frac{\pi}{2 n}} 
\int_0^\infty d\xi e^{-\mathfrak{k} \xi^n + e^{-\imath \frac{\pi (n-1)}{2 n}} \mathfrak{l} \xi}
\right]
&
\quad\mbox{if $n$ is odd} 
\label{eq:KernelDefIa}\\
2 e^{\imath \frac{\pi}{2 n}} 
\int_0^\infty d\xi e^{-\mathfrak{k} \xi^n} 
\cos\left( \cos(\pi \frac{(n-1)}{2 n}) \mathfrak{l} \xi\right)
&
\quad\mbox{if $n$ is even} 
\end{array}
\right.
\label{eq:KernelDefIIa}
\end{eqnarray}
and $\mathfrak{k} := k \sigma^n$.
For small values of $k$ the conditional $n$-marginal pdf reads:
\begin{equation}
\tilde{\nu}^{(n)}(k | \delta t) =
\tilde{\varpi}
\left(
(\delta t)^{\beta \underline{\underline{E}}^{T}} 
\left(
\mathfrak{G} k^{1/n}
\vec{\sigma}
\right)
\right)
\label{eq:SmallKValNMargPdf}
\end{equation}
where 
\begin{equation}
\mathfrak{G} := \frac{\mathfrak{C} \pi }{2 \cos(\pi \frac{(n-1)}{2 n}) }
\label{eq:DefG}
\end{equation}
and $\mathfrak{C}$ is defined in Appendix A
The proof of the proposition is in Appendix A.
\begin{corollary}
The mean value of the $n^{\mbox{th}}$ power
of the fluctuation term in (\ref{eq:StockPriceFluct})
conditioned on the waiting time $\delta T = \delta t$
is infinite if $n$ is even and $n \ge 2$ and is zero if $n$ is odd.
\begin{equation}
\mbox{E}
\left[
\left(\vec{\sigma} \cdot \delta \vec{L}_t \right)^n \left| \;\delta T = \delta t \right.\right]
=
\left\{
\begin{array}{ll}
\infty    & \quad\mbox{for }\; n \in 2\mathbb{N} \quad\mbox{and $n \ge 2$}\\
0         & \quad\mbox{for }\; n \in \mathbb{N} \diagdown 2\mathbb{N} 
\end{array}
\right.
\label{eq:InfiniteMean}
\end{equation}
The first statement follows from 
$\left. \partial \tilde{\nu}^{(n)}(k | \delta t) / \partial k \right|_{k = 0} = \infty$
(see (\ref{eq:SmallKValNMargPdf})) 
and the second statement follows from the assumption that  
the conditional jump pdf $\phi(\vec{x}|\delta t)$ and hence the jump intensity $\varpi(\vec{x})$ are symmetric in $\vec{x}$ (see beginning of section \ref{sec:TheStockPrice}).
\end{corollary}
\begin{corollary}
The $n^{\mbox{th}}$ power
of the fluctuation term in (\ref{eq:StockPriceFluct}) has a form:
\begin{equation}
Z := \left(\vec{\sigma} \cdot \delta \vec{L}_t \right)^n = 
\mathfrak{R} (\delta t)^\beta + O\left( (\delta t)^{2\beta} \right)
\label{eq:NMarginalTimeScal}
\end{equation}
where $\mathfrak{R}$ is a random number that depends 
neither on time $t$ nor on $\delta t$. 

We have:
\begin{eqnarray}
\lefteqn{
\frac{1}{dz} P\left( z \le Z \le z + dz \left| \; \delta T = \delta t \right. \right) = 
\nu^{(n)}(z | \delta t) = \mathcal{F}^{-1}_{k} \left[ \tilde{\nu}^{(n)} \right] (z) =}
\label{eq:NMarginalTimeScalI}\\
&&
 \mathcal{F}^{-1}_{k} \left[ \tilde{\varpi}
\left(
(\delta t)^{\beta \underline{\underline{E}}^{T}} 
\left(
\mathfrak{G} k^{1/n}
\vec{\sigma}
\right)
\right)
\right] (z) =
 \mathcal{F}^{-1}_{k} \left[ 
\tilde{\varpi}^{(\delta t)^\beta}
\left(
\mathfrak{G} k^{1/n}
\vec{\sigma}
\right) 
\right](z)=
\label{eq:NMarginalTimeScalII}\\
&&
\mathcal{F}^{-1}_{k} \left[ 
1 - (\delta t)^\beta (\tilde{\varpi} - 1) 
\left(
\mathfrak{G} k^{1/n}
\vec{\sigma}
\right) 
+ O\left((\delta t)^{2 \beta}\right)
\right](z)=
\label{eq:NMarginalTimeScalIII}\\
&&
\delta(z - z_0) - 
(\delta t)^\beta
\mathcal{F}^{-1}_{k} \left[ 
(\tilde{\varpi} - 1) 
\left(
\mathfrak{G} k^{1/n}
\vec{\sigma}
\right) 
\right](z)
+ O\left((\delta t)^{2 \beta}\right)
\label{eq:NMarginalTimeScalIV}
\end{eqnarray}
for some $z_0 \in \mathbb{R}$.
(\ref{eq:NMarginalTimeScalIV}) is equivalent to (\ref{eq:NMarginalTimeScal})
for $z > z_0$. The random variable $\mathfrak{R}$ conforms to a distribution
whose characteristic function is 
$(\tilde{\varpi} - 1) 
\left(
\mathfrak{G} k^{1/n}
\vec{\sigma}
\right)$.
\end{corollary}

The concrete dependence of $\nu^{(n)}(k | \delta t)$ on time
$\delta t$ depends on the form 
of the matrix $ \underline{\underline{ E}}$ (stable index).
We investigate that dependence in next sections.
\end{proposition}

Now we are going to investigate the functional form of the jump intensity $\varpi(\vec{x})$
in the limit of large jumps $\vec{x}$.
It is more convenient to analyse the Fourier transform
$\tilde{\varpi}(\vec{k})$ for small values of $\vec{k}$.
This function is a solution 
of equation (\ref{eq:ScalingLimitCorFourTransf}) and is determined
in a unique way via the stable index
$\underline{\underline{E}} \in \mathbb{L}(\mathbb{R}^d)$.
Since, however, on the grounds of the Jordan decomposition theorem,
every matrix $\underline{\underline{E}}$ has a unique representation 
as a block diagonal matrix where every block is of the form:
\begin{equation}
\left(
      \begin{array}{rrrrr}
                    \mathfrak{a}\;\; & 0\;\; & 0\;\; & \dots\;\; & 0\;\;      \\
                    1\;\; & \mathfrak{a}\;\; & 0\;\; & \dots\;\; & 0\;\;      \\
                    0\;\; & 1\;\; & \mathfrak{a}\;\; & \dots\;\; & \vdots\;\; \\
                    \vdots\;\; & & \ddots\;\; & \ddots\;\; \\
                    0\;\; & \vdots\;\; & & 1\;\; & \mathfrak{a}\;\;  
      \end{array}
\right)
\quad\mbox{or}\quad
\left(
      \begin{array}{rrrrr}
                    \mathfrak{B}\;\; & 0\;\; & 0\;\; & \dots\;\; & 0\;\;      \\
                    I\;\; & \mathfrak{B}\;\; & 0\;\; & \dots\;\; & 0\;\;      \\
                    0\;\; & I\;\; & \mathfrak{B}\;\; & \dots\;\; & \vdots\;\; \\
                    \vdots\;\; & & \ddots\;\; & \ddots\;\; \\
                    0\;\; & \vdots\;\; & & I\;\;  & \mathfrak{B}\;\;  
      \end{array}
\right)
\end{equation}
where $\mathfrak{a}$ is a real eigenvalue of $\underline{\underline{E}}$ in the first case,
and in the second case
\begin{equation}
\mathfrak{B} = \left(\begin{array}{rr} \mathfrak{a} & -\mathfrak{b} \\ \mathfrak{b} & \mathfrak{a} \end{array}\right)
\quad\mbox{and}\quad
I = \left(\begin{array}{rr} 1 & 0 \\ 0 & 1 \end{array}\right)
\end{equation} 
where $\mathfrak{a} \pm \imath \mathfrak{b}$ 
is a complex conjugate pair of eigenvalues of $ \underline{\underline{ E}}$
the set of all possible jump intensities $\varpi$ is narrowed down 
to few classes of solutions only, each one corresponding to a particular 
Jordan decomposition of the matrix  $\underline{\underline{E}}$.
In the following we investigate these classes of solutions
as a function of $\underline{\underline{E}}$.
We proceed as follows:
\begin{enumerate}
\item Take a Jordan decomposition of $\underline{\underline{E}} \in \mathbb{L}(\mathbb{R}^d)$,
\item Find  the Fourier transform $\tilde{\varpi}(\vec{k})$ of the jump intensity 
       by solving the functional equation 
      (\ref{eq:ScalingLimitCorFourTransf})
\item Find, from (\ref{eq:FourTrNMarginalProp})
      the Fourier transform $\tilde{\nu}^{(n)}(k | \delta t)$ 
      of the conditional $n$-marginal pdf (see Table \ref{tab:SummaryDef})
      as a function $\delta t$ and of $\vec{k}$ for small $\delta t$ and small $\vec{k}$.
\end{enumerate}
In the following sections we seek for the solutions $\varpi$ of the equation (\ref{eq:ISExpDef})
for different types of blocks in the Jordan representation.
The names of the sections refer to the type of Jordan decompositions of
$\underline{\underline{E}}$.

\subsection{The pure scaling\label{sec:Scal}}
Let $\underline{\underline{E}} = (d\mu)^{-1} I$ where $I$ is a 
$d$ dimensional identity matrix and $\mu > 0$ is a constant.
Here $\mbox{Tr$\underline{\underline{E}} = \mu^{-1}$}$.
The mapping:
\begin{equation}
t^{\underline{\underline{E}}^{T}} : 
\mathbb{R}^d \ni \vec{k}
\rightarrow t^{(d\mu)^{-1}} \vec{k}  \in \mathbb{R}^d
\end{equation} 
changes the length of $\vec{k}$ only.
We therefore assume the jump intensity $\varpi(\vec{k}) = \varpi(|\vec{k}|)$ 
to depend on the length of the vector only.
Substituting that assumption
into (\ref{eq:ScalingLimitCorFourTransf}) 
and taking $t^{(d\mu)^{-1}} = |\vec{k}|^{-1}$ we get:

\begin{equation}
 {\tilde \varpi}(|\vec{k}|)^{ |\vec{k}|^{-(d\mu)} } =
{\tilde \varpi}(|\vec{k}|^{-1} |\vec{k}|) 
\end{equation}
and
\begin{equation}
{\tilde \varpi}(\vec{k}) = {\tilde \varpi}(1)^{|\vec{k}|^{d \mu}} =
\exp\left(  \mathcal{C}|\vec{k}|^{d \mu} \right)
\label{eq:PureScalRes}
\end{equation}
where $\mathcal{C} :=\log({\tilde \varpi}(1))$. 

The conditional $n$-marginal pdf (see Table \ref{tab:SummaryDef})
from (\ref{eq:FourTrNMarginalProp}) reads:
\begin{eqnarray}
\tilde{\nu}^{(n)}(k|\delta t) = 
\int_{-\infty}^\infty d\mathfrak{l}
\exp\left\{
\mathcal{C} (\delta t)^\beta |\mathfrak{l}|^{d \mu}
\right\}
\mathcal{K}^{(n)}(k \sigma^n, \mathfrak{l}) 
\end{eqnarray}
where the kernel $\mathcal{K}$ is defined in 
(\ref{eq:KernelDefII}) and has a series expansion in $\mathfrak{l}$
given in (\ref{eq:KernelExpansionIV}).
From (\ref{eq:SmallKValNMargPdf}) we get the $n$-marginal pdf for $n$ even and for small values of $k$.
We have:
\begin{eqnarray}
\tilde{\nu}^{(n)}(k|\delta t) 
&=& \exp\left\{
\mathcal{C} 
\left( \frac{\mathfrak{C} \sigma \pi}{2 a_n }\right)^{d \mu}
(\delta t)^\beta
k^{\frac{d\mu}{n}}
\right\}
\end{eqnarray}
where $a_n := \cos(\pi \frac{(n-1)}{2 n})$ and the constant 
$\mathfrak{C}$ is defined in Appendix A.

\subsection{The scaling \& rotation\label{sec:ScalRot}}
Take $d=2$ and let:
\begin{equation}
\underline{\underline{E}} = 
\left( 
\begin{array}{cc}
(2\mu)^{-1} & -b          \\
b           & (2\mu)^{-1} 
\end{array}
\right)
\end{equation}
The trace $\mbox{Tr$\left[\underline{\underline{E}}\right] = \mu^{-1}$}$.
We denote by 
$\underline{\underline{ O}}_{\beta} := 
\left( 
\begin{array}{rr} \cos(\beta) & -\sin(\beta) \\
                  \sin(\beta) & \cos(\beta)
\end{array}
\right)$ a two dimensional rotation by an angle $\beta$. The mapping:
\begin{equation}
t^{\underline{\underline{E}}^{T}} : 
\mathbb{R}^2 \ni \vec{k}
\rightarrow 
t^{(2 \mu)^{-1}} 
\underline{\underline{ O}}_{-b \log(t)} \vec{k}
\in \mathbb{R}^2
\end{equation}
changes the length of $\vec{k}$ by a factor $t^{(2 \mu)^{-1}}$
rotates by an angle $-b \log(t)$.

We seek solutions of (\ref{eq:ScalingLimitCorFourTransf}) in the form:
\begin{equation}
\tilde{\varpi}(\vec{k}) = \exp\left( \tilde{\upsilon}(|\vec{k}|,\theta) \right)
\label{eq:OpStableSolAnsatz}
\end{equation}
where the function $\tilde{\upsilon}(\vec{k}) = \tilde{\upsilon}(|\vec{k}|,\theta)$ is assumed to depend on the length $|\vec{k}|$
and on the angle \mbox{$\theta := \sphericalangle(\vec{k},(1,0))$}.
Inserting (\ref{eq:OpStableSolAnsatz}) into 
(\ref{eq:ScalingLimitCorFourTransf}) we obtain:
\begin{equation}
t\tilde{\upsilon}(|\vec{k}|,\theta) = \tilde{\upsilon}(t^{(2\mu)^{-1}}|\vec{k}|, \theta  - b \log(t))
\label{eq:Eq4Upsilon}
\end{equation}
Now we assume that the function $\tilde{\upsilon}$ factorizes:
\begin{equation}
\tilde{\upsilon}(\vec{k}) = \tilde{\upsilon}_1(|\vec{k}|) \Theta(\theta)
\label{eq:ScalRotAnsatz}
\end{equation}
Inserting (\ref{eq:ScalRotAnsatz}) into (\ref{eq:Eq4Upsilon}) 
and taking $t^{(2 \mu)^{-1}} = |\vec{k}|^m$ with $m\in \mathbb{N}$ we get:

\begin{equation}
|\vec{k}|^{2\mu m} \tilde{\upsilon}_1(|\vec{k}|) \Theta(\theta) =
\tilde{\upsilon}_1(|\vec{k}|^m |\vec{k}|) \Theta(\theta - b \log(t)) =
\tilde{\upsilon}_1(|\vec{k}|^{m+1}) \Theta(\theta - 2\mu b m \log(k))
\end{equation}
After separating variables and assuming that the jump intensity $\varpi(\vec{x})$ is even in $\vec{x}$ we obtain $\tilde{\upsilon}$
in the following form: 
\begin{equation}
\tilde{\upsilon}(\vec{k}) = \tilde{\upsilon}_1(1) k^{2\mu} 
\sum_{q=-\infty}^\infty
\mbox{Re}
\left[ 
 \exp\left[\imath \frac{\pi}{\mu b} \frac{n_q}{m_q} \frac{\theta}{\log(k)} 
     \right] 
\right]
\label{eq:FourJumIntSolIIa}
\end{equation}
where $(n_q,m_q) \in \mathbb{N}$ for $q\in\mathbb{Z}$
are some numbers that depend on $t$.
We find only solutions for $t = |\vec{k}|^{2\mu m}$ for $m \in \mathbb{N}$
and leave the generic case for future work.
We check that (\ref{eq:FourJumIntSolIIa}) is indeed a solution of (\ref{eq:Eq4Upsilon})
for $t = |\vec{k}|^{2\mu m}$ under following assumptions:
\begin{equation}
\mathop{\forall}_{q\ge 1}
\quad
\mathop{\exists}_{p' \in \mathbb{N}}
\quad
\frac{n_q}{m_q (m + 1)} =  \frac{n_{p'}}{m_{p'}}
\quad\mbox{and}\quad
\frac{n_q}{m_q} \frac{m}{m + 1} = 2 \mathbb{N}
\end{equation}
This leads to a following solution for the jump intensity in the Fourier domain:
\begin{equation}
\tilde{\varpi}(\vec{k}) = 
\exp\left[
\mathcal{D} k^{2\mu} 
\sum_{q=-\infty}^\infty
\cos\left(
 \frac{2\pi}{\mu b} \frac{(m + 1)^{|q| + 1}}{m} \frac{\theta}{\log(k)} 
\right)
\right]
\label{eq:FourJumIntSolIIb}
\end{equation}
where $\mathcal{D} := \tilde{\upsilon}_1(1)$
For small $k$ the conditional $n$-marginal pdf reads:
\begin{equation}
\tilde{\nu}^{(n)}(k|\delta t)  = 
\exp\left[ 
\mathcal{D} 
\left( \mathfrak{G} \sigma k^{1/n} \right)^{2 \mu}
(\delta t)^\beta 
\sum_{q=-\infty}^\infty
\cos\left[  
 \frac{2\pi}{\mu b} 
 \frac{(m + 1)^{|q| + 1}}{m} 
 \Omega(\delta t, \sigma, k) 
\right]
\right]
\end{equation}
where 
\begin{equation}
\Omega(\delta t, \sigma, k) := 
\left(
\frac{\theta  - b \beta \log(\delta t)}
     {\log(\mathfrak{G} \sigma k^{1/n}) + \frac{\beta}{2\mu} \log(\delta t)}
\right)
\end{equation}
where the number $\mathfrak{G}$ is defined in (\ref{eq:DefG}).
We summarise the results for the Fourier transforms of the jump intensity 
and of the conditional $n$-marginal pdf in both cases of pure scaling
and of scaling \& rotation: 
\begin{center}
\fbox{\begin{minipage}{\textwidth}
\begin{eqnarray}
{\tilde \varpi}(\vec{k}) &=& 
\left\{
\begin{array}{ll}
\exp\left(  \mathcal{C}|\vec{k}|^{d \mu} \right)
& \mbox{pure scaling} \\
\exp\left[
\mathcal{D} k^{2\mu} 
\sum\limits_{q=-\infty}^\infty
\cos\left(  
 \frac{2\pi}{\mu b} \frac{(m + 1)^{|q| + 1}}{m} \frac{\theta}{\log(k)} 
\right)
\right]
 & \mbox{scaling \& rotation}
\end{array}
\right.
\\
\tilde{\nu}^{(n)}(k|\delta t)  &=& 
\left\{
\begin{array}{ll}
\exp\left\{
\mathcal{C} 
\left( \mathfrak{G} \sigma k^{1/n} \right)^{d \mu}
(\delta t)^\beta
\right\}
& \mbox{pure scaling} \\
\exp\left[ 
\mathcal{D} 
\left( \mathfrak{G} \sigma k^{1/n} \right)^{2 \mu}
(\delta t)^\beta 
\sum\limits_{q=-\infty}^\infty
\cos\left[  
 \frac{2\pi}{\mu b} 
 \frac{(m + 1)^{|q| + 1}}{m} 
 \Omega(\delta t, \sigma, k) 
\right]
\right]
& \mbox{scaling \& rotation} \\
\end{array}
\right.
\label{eq:SummaryofResults}
\end{eqnarray}
\end{minipage}
} 
\end{center}
where $\mathfrak{G}$ is defined in (\ref{eq:DefG})
and the results for the conditional $n$-marginals hold for small $k$ only.

\subsection{The joint transition probability\label{eq:JointPdfs}}
The purpose of this section is to derive probability distributions
of financial instruments like barrier European style options
with multiple exercise times and American style options (exotic options) \cite{MusielaRutkowski} 
as a function of of the conditional $1$-marginal pdf. 
We derive a joint probability distribution of log-prices
$X_t = \log(S_t)$ of the stock
at certain times and of the maximal log-price in some time interval.

We fix $l+1$ values of time $t_0 < t_1 < t_2 < \dots < t_l $ 
and we analyse the joint cumulative distribution 
function $F_{X_{t_1},X_{t_2},\dots,X_{t_l};M_{t_l}}(x_1,x_2,\dots,x_l;y)$
related to the log-prices $X_{t_i}$ for $i=1,\dots,l$ of the stock
and to the maximal value of the log-price $M_t := \mathop{\mbox{sup}}_{0 \le s \le t} X_s$.
We set $X_{t_0} = 0$, we define $\delta t = (t_l - t_0)/N$ and we define
integers $1 \le i_1 < \dots < i_l \le N$ such that 
$t_j - t_0 = (\delta t) i_j$. From the definition of the joint cumulative distribution function
and from the the SDE (\ref{eq:StockPriceFluct}) we have:
\begin{eqnarray}
\lefteqn{
F_{X_{t_1},X_{t_2},\dots,X_{t_l};M_{t_l}}(x_1,x_2,\dots,x_l;y) := 
P\left( X_{t_1} < x_1, \dots, X_{t_l} < x_l; M_{t_l} < y\right) 
\label{eq:JointWalkerMaxPdfI} } \\
&&= P\left( \bigcup \limits_{j=1}^l X_{t_0 + i_j (\delta t)} < x_j;
            \bigcup \limits_{j=1}^N X_{t_0 + j (\delta t)} < y 
     \right) 
\label{eq:JointWalkerMaxPdfII}\\
&&= P\left( \bigcup \limits_{j=1}^l 
            \left( \sum_{p=1}^{i_j} \vec{\sigma} \cdot \vec{\delta L}_{t_0 + p \delta t} \right)
            < (x_j - t_j \mu);
            \bigcup \limits_{j=1}^N 
            \left( \sum_{p=1}^j \vec{\sigma} \cdot \vec{\delta L}_{t_0 + p \delta t} \right)
            < (y - (t_0 + j \delta t) \mu)
     \right)
\nonumber \\ %
\label{eq:JointWalkerMaxPdfIII}\\
&&= \int\limits_{\mathbb{R}^n} 1_{\Delta (x_1,\dots,x_l;y)} \prod\limits_{j=1}^N \nu (\xi_j | \delta t) d\xi_j
\label{eq:JointWalkerMaxPdfIV}
\end{eqnarray}
where $\bigcup \limits_{j=1}^l $ denotes a union of $l$ conditions, the integration region 
\begin{equation}
\Delta (x_1,\dots,x_l;y) := \left\{ 
\begin{array}{l}
\bigcup \limits_{j=1}^l 
\left( \sum_{p=1}^{i_j} \xi_p \right) < (x_j - t_j \mu) \\
\bigcup \limits_{j=1}^N 
            \left( \sum_{p=1}^j \xi_p \right) < (y - (t_0 + j \delta t) \mu )
\end{array}
\right.
\label{eq:InegrationRegion}
\end{equation}
is an infinite region in $\mathbb{R}^N$ bounded by $l+N$ hyperplanes
of dimension $N-1$ and $\nu(\xi | t) := \nu^{(1)}_{\vec{\sigma} \cdot \vec{\delta L}}(\xi | t)$ 
is the conditional probability density of the random variable $\vec{\sigma} \cdot \vec{\delta L}$.
The Fourier transform of that probability density is given in (\ref{eq:SummaryofResults}).
Now the procedure consists of following three steps.
\begin{itemize}
\item[(a)] Derive the joint probability density function 
$f_{X_{t_1},X_{t_2},\dots,X_{t_l};M_{t_l}}(x_1,x_2,\dots,x_l;y)$
by differentiating
the cumulative density function,
\item[(b)] Derive the joint characteristic function by taking Fourier transforms
of all $l+1$ random variables and expressing the result through Fourier transforms
$\tilde{\nu}^{(1)}(k| \delta t) := \mathcal{F}_z\left[\nu^{(1)}(z | \delta t)\right](k)$
of the pdf of the random variable $\vec{\sigma} \cdot \vec{\delta L}$
\item[(c)] Derive the joint pdf in (a) by inverting expressions 
containing the Fourier transforms $\tilde{\nu}^{(1)}(k| \delta t)$ in obtained (b).
\end{itemize}
We describe the steps in the following and give examples.
\begin{itemize}
\item[(a)]

\begin{eqnarray}
\lefteqn{
f_{X_{t_1},X_{t_2},\dots,X_{t_l};M_{t_l}}(x_1,x_2,\dots,x_l;y) := }
\\
&&\frac{\partial^{l+1}}{\partial x_1 \dots \partial x_l \partial y}
\left[  F_{X_{t_1},X_{t_2},\dots,X_{t_l};M_{t_l}}(x_1,x_2,\dots,x_l;y) \right] 
\label{eq:JointPdfI} \\
&&
\!\!\!\!\!\!\!\!\!\!\!\!\!\!\!\!\!\!\!\!\!\!\!\!\!\!\!\!\!\!
\sum_{j=1}^l \sum_{i=1}^N
\int\limits_{\mathbb{R}^N} 
\delta\left( x_j - t_j \mu - \sum_{p=1}^{i_j} \xi_p \right) 
\delta\left( y - (t_0 + i \delta t) \mu - \sum_{p=1}^i \xi_p \right) 
\prod_{j=1}^N \nu(\xi_j|\delta t) d\xi_j
\label{eq:JointPdfII}
\end{eqnarray}
The sum on the right-hand side in (\ref{eq:JointPdfII}) corresponds to intersections
of all possible pairs of hyperplanes that bound the region
(\ref{eq:InegrationRegion}).
\item[(b)]
The joint characteristic function of the $l+1$ random variables in question reads:
\begin{eqnarray}
\lefteqn{
\chi_{X_{t_1},X_{t_2},\dots,X_{t_l};M_{t_l}}(k_1,k_2,\dots,k_l;w) :=}
\\
&&E\left[ \exp(\imath \left(k_1 X_{t_1} + \dots + k_l X_{t_l} + w M_{t_l}\right)) \right]
\label{eq:JointChiI} \\
&&\frac{1}{l N} \sum_{j=1}^l \sum_{i=1}^N 
\left\{ \begin{array}{ll}
        (\tilde{\nu}(k_j + w | \delta t))^i (\tilde{\nu}(k_j | \delta t))^{i_j - i} & \mbox{  if $i \le i_j$} \\
        (\tilde{\nu}(k_j + w | \delta t))^{i_j} (\tilde{\nu}(w | \delta t))^{i-i_j} & \mbox{  otherwise}
\end{array}
\right\}
\label{eq:JointChiII} 
\end{eqnarray}
where in  
(\ref{eq:JointChiII}) we replaced the delta functions in the integrand by exponentials 
$\exp(\imath k_j x_j)$ and $\exp(\imath w y)$
and we we factored out expressions depending on $\xi_p$.

{\bf Example:}
\begin{enumerate}
\item We take $l=1$ and we obtain the joint characteristic function of the log-price
and of the maximum of the log-price. We have:
\begin{eqnarray}
E\left[ \exp(\imath \left(k X_t + w M_t \right)) \right] &=& 
\frac{1}{N} (\tilde{\nu}(k | \delta t))^{i_1}
\sum_{i=1}^N \left(\frac{\tilde{\nu}(k + w | \delta t)}{\tilde{\nu}(k | \delta t)}\right)^i
\end{eqnarray}
\end{enumerate}

\end{itemize}

The result (\ref{eq:JointChiII}) will be used in in future work for pricing exotic options
and for deriving variants of the Wiener-Hopf factorisation
formula \cite{SatoLevyProcess} for the operator stable L\'{e}vy process in question.

\subsection{The option price \label{sec:OptionPrice}}
An option on a financial asset is an agreement settled at time $t$
to purchase (call) or to sell (put)
the asset at some time $T$ (maturity) in the future.
For the sake of simplicity we consider European style options, 
ie such that can be only exercised 
at maturity 
(this means that boundary conditions are imposed on the option price at maturity $t=T$).
Extending the analysis to American style options 
(to be exercised at any time) is possible and can be done by considering
European style options with $m$ different maturities \cite{Dash1}  
and performing the limit $m\rightarrow \infty$.

In order to minimise the risk we diversify the 
portfolio by dividing the money available in to $N_S$ stocks $S_t$
and $N_C$ options $C(S_t;t)$.
This means that the portfolio $V(t)$ reads:
\begin{equation}
V(t) = N_S S_t + N_C C(S_t;t)
\label{eq:portfolioDef}
\end{equation}
where the coefficients $N_S$ and $N_C$ are customarily chosen \cite{MusielaRutkowski}
as $N_S = -\partial C /\partial S N_C$ and in the following we take $N_C = 1$.

The investment strategy consists in replicating the portfolio, ie
investing a copy of it into a safe central-bank account and let it earn
a compound interest at a rate $r$ independent of $t$.
We analyse the distribution of 
deviations 
\begin{equation}
\mathfrak{D_t} := V(t + \delta t) - e^{r (\delta t)^\beta} V(t)
\label{eq:DevDefinition}
\end{equation}
between the compound interest $e^{r (\delta t)^\beta} V(t) - V(t)$ 
earned in the safe central-bank account and changes $V(t + \delta t)- V(t)$ 
of the price of the portfolio and we check if a self-financing strategy exists,
ie if it is possible to choose $C = C(S_t;t)$ subject to a condition 
$C_T = \mbox{max$\left( S_T - K, 0 \right)$}$ such that the average
value of the portfolio deviation equals zero 
\begin{equation}
E\left[ \mathfrak{D_t} \right] = 0
\label{eq:NoDrift}
\end{equation} 
(the deviations have no drift)
and the variance of this deviation 
$\mbox{Var$\left[ \mathfrak{D_t} \right]$}$
is minimal. 
Since we assumed (\ref{eq:ScalinglimitTh}) 
that the waiting times between stock price changes
are power law distributed with index $\beta$  
we need to take
the infinitesimal interest earned in time period $\left[t,t+\delta t\right]$
equal to $r (\delta t)^\beta$ in order to ensure that the
deviation variable
$\mathfrak{D_t}$  exists in the limit $\delta t \rightarrow 0$.

Our approach differs here from the approach \cite{RamaCont} in financial mathematics 
where one assumes in the first place that the deviations $\mathfrak{D_t}$ have no drift 
(there are no arbitrage opportunities on the market) and 
one imposes conditions (semi-martingale condition) on the high end of the jump intensity 
$\varpi(\vec{x}) \sim e^{-|\vec{x}|}$ in order to ensure this assumption.
We cannot make such an assumption for operator stable L\'{e}vy processes.
The processes that we are using are sums of a large number of of 
independent, identically distributed jumps where the distribution of the latter 
diminish according to a power law
and not to an exponential in the high end.
The processes are invariant under auto-convolutions (\ref{eq:ISExpDef})
(self-similar) and in our opinion it is much more appealing from the physical
point of view 
\footnote{We understand the price formation process as an accumulation 
of a large number of shocks that, once they exceed some threshold value, 
give rise to a change of the stock price 
(see discussion under Table \ref{tab:SummaryDef})}
to use them in 
financial modelling rather than to use the abstract notion of 
the metric space of 
L\'{e}vy processes
with right continuous and left bounded paths endowed  with a Shorokhod 
topology if there is no experimental justification that financial time series in question are martingales.

Our approach is not equivalent with theory in chapters {\bf 20.4.3} -- {\bf 20.4.5}
pages 1416 -- 1428 in \cite{PathIntegrals}.
That theory assumes that the log-characteristic function of the fluctuations
that drive $\log S_t$ is analytic (is power expandable)
and derives a partial differential equation (Fokker-Planck equation)
for the price of the option as a function of the price of the stock and of time
and solves this equation subject to initial conditions for the price of the stock 
at maturity.
Operator stable processes are not power expandable.
It might be possible to expand the log-characteristic function
$\tilde{\varpi}(\vec{k})$ in a series of fractional powers of $\left|\vec{k}\right|$
using the fractional calculus \cite{Samko}
and in particular the fractional Taylor expansion
(formula (4.11) page 89 in \cite{Samko}) \cite{Dzherbashyan}
and develop a fractional Fokker-Planck equation for the price of the option on the stock.
In future work we will pursue investigations in this context
in order to reconcile our results with that from \cite{PathIntegrals}.

Equation (\ref{eq:StockPriceFluct}) defines the price of a stock
in terms of an exponential from a L\'{e}vy process.
Thus a small change of the stock price reads: 
\begin{eqnarray}
S_{t + \delta t} - S_t &=& 
S_t \left(
e^{\alpha (\delta t)^\beta} \exp\left[ 
                 \left(\vec{\sigma} \cdot \delta \vec{L}_t\right) 
                 \right] -1
\right) 
\label{eq:StockPriceChangeI}
\\
&=&
S_t \left(
\alpha (\delta t)^\beta + \sum_{m=1}^{\infty} 
\frac{ \left(\vec{\sigma} \cdot \delta \vec{L}_t\right)^m}{m!} 
\right)
\label{eq:StockPriceChangeII}
\end{eqnarray}
where in (\ref{eq:StockPriceChangeII}) 
we expanded the exponential in a Taylor series and
retained all terms of the expansion since all these terms are proportional to 
$(\delta t)^\beta$ (\ref{eq:NMarginalTimeScal})
for large values of $\delta \vec{L}_t$.

From (\ref{eq:StockPriceChangeI}) and (\ref{eq:StockPriceChangeII})
we see that the price of the stock $S_t$ is not a martingale.
It is also not possible to construct a martingale from the price of 
the stock by subtracting the sum of means 
$E\left[ S_{t+\delta t} - S_t  \left| S_t \right. \right]$ over $t$ (the compensator)
from $S_t$ \cite{EberleinOezkan}.
Indeed, the mean value of the left hand side of  (\ref{eq:StockPriceChangeI})
conditioned on the history of the stock price up to time $t$ is 
a sum of all possible positive powers of the fluctuation term 
$\left(\vec{\sigma} \cdot \delta \vec{L}_t\right)^m$ and is,
on the grounds of (\ref{eq:InfiniteMean}), infinite.
It is therefore readily seen that, in the framework of our model,
satisfying the condition  (\ref{eq:NoDrift})
and thus solving the option pricing problem in its classical formulation
may not be possible.
We will therefore analyse the probability distribution of the deviation variable
$\mathfrak{D_t}$ and work out conditions under which it is possible to minimise
the drift $E\left[\mathfrak{D_t}\right]$ of the portfolio (\ref{eq:portfolioDef}).

We expand a change of the price of an option $C(S_t;t)$ on the stock
in a Taylor series in powers of the change of the stock price:
\begin{equation}
C_{t+\delta t} - C_t =
\frac{\partial C}{\partial t} (\delta t)^\beta +
\frac{\partial C}{\partial S} (S_{t+\delta t} - S_t) + 
\sum_{n=2}^\infty 
\frac{\partial^n C}{\partial S^n} \frac{(S_{t+\delta t} - S_t)^n}{n!}
\label{eq:OptionChange}
\end{equation}
where the partial derivatives 
$\partial C/\partial t$ an $\partial^n C/\partial S^n$
are assumed to be independent of time $t$ and they will be determined
later.

From (\ref{eq:StockPriceChangeI}) and (\ref{eq:StockPriceChangeII})
 we compute the $n$th powers of the changes of the stock price
and we retain only those terms that are proportional to $(\delta t)^\beta$.
We have 
\begin{equation}
\left(S_{t+\delta t} - S_t\right)^n = S_t^n
\left( 
\exp\left[ 
                 \left(\vec{\sigma} \cdot \delta \vec{L}_t\right) 
                 \right] 
-1
\right)^n
\quad\mbox{for $n \ge 2$} 
\label{eq:StockNChange}
\end{equation}
where we have neglected all terms 
$\left(\mu \delta t\right)^{n_1} \left(\vec{\sigma}\cdot \delta \vec{L}_t\right)^{n_2}$
on the grounds of (\ref{eq:NMarginalTimeScal}).
We insert (\ref{eq:OptionChange}) and (\ref{eq:StockNChange})
into (\ref{eq:DevDefinition}) and we get the deviation random variable:
\begin{equation}
\mathfrak{D_t} = 
\left( \partial_t C + r S_t \partial_S C - r C_t \right) (\delta t)^\beta +
\sum_{n=2}^\infty \frac{1}{n!} \frac{\partial^n C}{\partial S^n} S_t^n \left( \exp\left(\vec{\sigma} \cdot \delta \vec{L}_t\right) - 1 \right)^n
\label{eq:DeviationResult}
\end{equation}
Note that all terms on the right hand side of (\ref{eq:DeviationResult})
are proportional to $(\delta t)^\beta$.
Now we derive the Fourier transform of the probability density function
of $\mathfrak{D}_t$ conditioned on the value of the price of the stock
at time $t$ (conditional deviation characteristic function). We have:
\begin{eqnarray}
\lefteqn{
\chi_{\mathfrak{D}_t\left|S_t\right.}(k) 
:= \mbox{E}\left[ e^{\imath k \mathfrak{D}_t} \left| S_t \right. \right] = }
\label{eq:CharacFctDevI}\\
&& 
\exp( \imath k D_t (\delta t)^\beta )
\int_{-\infty}^\infty dz 
\nu^{(1)}(z | \delta t)
\exp\left( \imath k
\sum_{n=2}^\infty \frac{1}{n!} 
\frac{\partial^n C}{\partial S^n} 
 S_t^n \left( \exp\left(z\right) - 1 \right)^n
\right) =
\label{eq:CharacFctDevII}\\
&&
\exp( \imath k D_t (\delta t)^\beta )
\int_{-\infty}^\infty d\lambda 
\tilde{\nu}^{(1)}(\lambda | \delta t)
\mathfrak{M}(k, \lambda) =
\label{eq:CharacFctDevIII}\\
&&
\exp( \imath k D_t (\delta t)^\beta )
\int_{-\infty}^\infty d\lambda 
\tilde{\varpi}((\delta t)^{\beta \underline{\underline{E}}^{T}} 
\frac{\vec{\sigma}}{|\vec{\sigma}|}(\lambda))
\mathfrak{M}(k, \lambda) 
\label{eq:CharacFctDevIV}
\end{eqnarray}
where \mbox{$D_t := \left( \partial_t C + r S_t \partial_S C - r C_t \right)$}
and
\begin{equation}
\mathfrak{M}(k, \lambda) := 
\frac{1}{2\pi}
\int_{-\infty}^\infty dz
\exp\left( \imath k
\sum_{n=2}^\infty \frac{1}{n!} 
\frac{\partial^n C}{\partial S^n} 
 S_t^n \left( \exp\left(z\right) - 1 \right)^n
\right) \exp(-\imath \lambda z) 
\end{equation}
and in (\ref{eq:CharacFctDevIV}) we made use of (\ref{eq:FourTrNMarginalProp})
and of (\ref{eq:KernelDefI}).
In further calculations we 
make use of the following proposition.
\begin{proposition}
The integral
\begin{equation}
\mathfrak{N}(k, \lambda) := 
\frac{1}{2\pi} \int_{-\infty}^\infty dz
\exp\left\{\imath k P(e^z - 1)\right\} e^{-\imath \lambda z}
\label{eq:PolDeltaFct}
\end{equation}
where $P(z)$ is a real function such that the only root of $P$ is $P(0) = 0$ 
and $P(\infty) = \infty$
tends, for small values of $k$, to a following delta function:
\begin{equation}
\mbox{lim}_{k \rightarrow 0} \mathfrak{N}(k, \lambda) = 
\delta\left(\lambda + 
\frac{\imath \mathfrak{m}^{-1} }
{\left( \log(P^{(-1,i)}\left(\frac{\imath \epsilon}{k}\right) \right)}\right)
%
\label{eq:LimLogDelta}
\end{equation}
where $\mathfrak{m}> 0$ is a number that satisfies 
(\ref{eq:DeltaConstEval}),
$P^{(-1)}(z)$ is the inverse function to $P(z)$ in some vicinity of $z=0$
and $\epsilon > 0$ is a positive real number.
The proof is given in \mbox{Appendix B}.
\end{proposition}
Inserting (\ref{eq:LimLogDelta}) into (\ref{eq:CharacFctDevIV}) we get
the conditional deviation characteristic function in the limit of small $k$.
We have:
\begin{equation}
\chi_{\mathfrak{D}_t\left|S_t\right.}(k) =
\exp( \imath k D_t (\delta t)^\beta )
\tilde{\varpi}
\left(
(\delta t)^{\beta \underline{\underline{E}}^{T}} 
\frac{\vec{\sigma}}{|\vec{\sigma}|}
( \frac{-\imath \mathfrak{m}^{-1} }
       {\left(\log\left(P^{(-1,i)}\left(\frac{\imath \epsilon}{k}\right)\right) \right)}) 
\right)
\end{equation}
where 
\begin{equation}
P\left(x\right) := 
\sum_{n=2}^\infty \frac{1}{n!} \frac{\partial^n C}{\partial S^n} S_t^n x^n 
\end{equation}
and we check by direct substitution that 
for $d=1$ and $\underline{\underline{E}} = \mu^{-1}$ the 
class of functions:
\begin{equation}
P(x) : = \sum_{n=2}^\infty \frac{1}{n!} \frac{\partial^n C}{\partial S^n} S_t^n x^n 
       \mathop{=}_{x\rightarrow \infty} \mathfrak{T} (\log(x))^{\alpha}
\label{eq:OptionPricingSol}
\end{equation}
with $0 < \alpha < \mu$ and $\mathfrak{T} \in \mathbb{R}$ and $P(0) = 0$
ensures that the drift of deviations 
\begin{equation}
\mbox{E}\left[ \mathfrak{D}_t \left| S_t \right. \right] = 
\left. \frac{\partial \chi_{\mathfrak{D}_t\left|S_t\right.}(k)}{\partial k} \right|_{k=0}
= 0
\label{eq:NoDriftI}
\end{equation} 
is zero.
The equation (\ref{eq:NoDriftI}) yields a differential equation 
for the option $C = C(S_t,t)$ as a function of the price of the stock $S_t$
and of time $t$ viz
\begin{eqnarray}
0 &=& D_t + 
\mbox{lim$_{k\rightarrow 0}$} ( \frac{\mathfrak{m}^{-\mu} }
{\imath k \left(\log\left(P^{(-1,i)}\left(\frac{\imath \epsilon}{k}\right)\right) \right)}) 
\label{eq:DifferentialEquationI} \\
&=& \partial_t C + r S_t \partial_S C - r C_t
 -\frac{1}{\epsilon \mathfrak{m}^\mu} 
\mbox{lim$_{k\rightarrow 0}$} 
\frac{
P^{(-1,i)}\left(\frac{\imath \epsilon}{k}\right)
}{
(P^{(-1,i)})^{'}\left(\frac{\imath \epsilon}{k}\right)
}
\label{eq:DifferentialEquationII} \\
&=& 
\left( 
\partial_t C + r S_t \partial_S C +\frac{S_t^2}{2} \partial_{S^2} C - r C_t
\right)
 -
\left(
\frac{1}{\epsilon \mathfrak{m}^\mu} 
\mbox{lim$_{k\rightarrow 0}$} 
\frac{
P^{(-1,i)}\left(\frac{\imath \epsilon}{k}\right)
}{
(P^{(-1,i)})^{'}\left(\frac{\imath \epsilon}{k}\right)
}
+ \frac{P^{''}(0)}{2}
\right)
\label{eq:DifferentialEquationIII}
\end{eqnarray}
Note that the first term enclosed by parentheses 
in (\ref{eq:DifferentialEquationIII}) 
corresponds to the classical Black\& Scholes equation
and the second term depends via (\ref{eq:OptionPricingSol})
on all partial derivatives $\partial_{S^n} C$.
In future work we will analyse the uniqueness 
of the solution of the option pricing
problem and we will derive conditions for the solution to have:
\begin{equation}
C_T = C(S_T,T) = \mbox{max$(S_T - K, 0)$}
\end{equation}
a given initial value at maturity $T$.

\section{Conclusions}
We have applied the technique of characteristic functions
to the problem of pricing an option on a stock that is driven
by operator stable fluctuations.
We have shown that it is possible to construct an option
on a stock as a function of the price of the stock in such a way
that deviations of the portfolio are compensated by the compound interest
earned on the portfolio. 
In future work we will price exotic options with different
exercise times by computing the unconditional, joint characteristic function
of the deviations of the portfolio related to values of stock price at these
exercise times and by using initial conditions on the value of the portfolio
at these exercise times.
Testing of the theory in Monte Carlo simulations
and fitting the theory to market data will also be pursued in future work.

\section{Appendix A}
We prove formula (\ref{eq:FourTrNMarginalProp}) for the conditional 
$n$-marginal probability function:
\begin{eqnarray}
\lefteqn{
\tilde{\nu}^{(n)}(k | \delta t) = 
\int_{-\infty}^\infty e^{\imath k z} \nu^{(n)}(z | \delta t) dz =}
\label{eq:NMarginal} \\
&&\int_{\mathbb{R}^d} \exp{\imath k \left(\vec{\sigma} \cdot \delta \vec{l}_t \right)^n}
\omega_{\vec{\delta L_t} | \delta T} \left( \delta \vec{l}_t| \delta t \right)
d( \delta \vec{l}_t ) =
\label{eq:NMarginalI} \\
&&\int_{\mathbb{R}^d} \exp{\imath k \left(\vec{\sigma} \cdot \delta \vec{l}_t \right)^n}
\phi \left( \delta \vec{l}_t| \delta t \right)
d( \delta \vec{l}_t ) =
\label{eq:NMarginalII} \\
&&(\delta t)^{-\beta \mbox{Tr$ \underline{\underline{ E}}$}}
\int_{\mathbb{R}^d} \exp{\imath k \left(\vec{\sigma} \cdot \delta \vec{l}_t \right)^n}
\varpi ((\delta t)^{- \beta \underline{\underline{E}}}\delta \vec{l}_t) 
d( \delta \vec{l}_t ) =
\label{eq:NMarginalIII} \\
&&
\!\!\!\!\!\!\!\!
\frac{(\delta t)^{-\beta \mbox{Tr$ \underline{\underline{ E}}$}}}{(2\pi)^d}
\int_{\mathbb{R}^d} d\vec{\lambda} \tilde{\varpi}(\vec{\lambda})
\int_{\mathbb{R}^d} \delta \vec{l}_t 
\exp(\imath \left( k \left(\vec{\sigma} \cdot \delta \vec{l}_t \right)^n - 
((\delta t)^{-\beta \underline{\underline{E}}^{T}} \vec{\lambda}) 
\cdot \delta \vec{l}_t \right)) =
\label{eq:NMarginalIV} \\
&&
\!\!\!\!\!\!\!\!\!\!
\frac{(\delta t)^{-\beta \mbox{Tr$ \underline{\underline{ E}}$}}}{(2\pi)^d}
\int_{\mathbb{R}^d} d\vec{\lambda} \tilde{\varpi}(\vec{\lambda})
\int_{\mathbb{R}^d} d(\delta \vec{l}_t)
\exp(\imath \left( \mathfrak{k} (\delta \vec{l}_t)_1^n - 
                   \vec{\mathfrak{l}} \cdot \delta \vec{l}_t
            \right) =
\label{eq:NMarginalIVa} \\
&&
\!\!\!\!\!\!\!\!\!\!
\frac{(\delta t)^{-\beta \mbox{Tr$ \underline{\underline{ E}}$}}}{(2\pi)^d}
\int_{\mathbb{R}^d} d\vec{\lambda} \tilde{\varpi}(\vec{\lambda})
(\prod_{i=2}^d 2 \pi \delta(\mathfrak{l}_i))
\int\limits_{-\infty}^\infty dw 
e^{\imath \mathfrak{k} (w^n - w)}
e^{\imath (\mathfrak{k} - \mathfrak{l_1}) w } =
\label{eq:NMarginalIVb} \\
&&
\!\!\!\!\!\!\!\!\!\!
\int\limits_{-\infty}^\infty d\mathfrak{l_1}
\tilde{\varpi}((\delta t)^{\beta \underline{\underline{E}}^{T}} 
\frac{\vec{\sigma}}{|\vec{\sigma}|}(\mathfrak{l_1}))
\frac{1}{2\pi} \int\limits_{-\infty}^\infty dw 
e^{\imath (\mathfrak{k} w^n - \mathfrak{l}_1 w)} =
\label{eq:NMarginalIVc} \\
&&
\!\!\!\!\!\!\!\!\!\!
\int\limits_{-\infty}^\infty d\mathfrak{l_1}
\tilde{\varpi}((\delta t)^{\beta \underline{\underline{E}}^{T}} 
\frac{\vec{\sigma}}{|\vec{\sigma}|}(\mathfrak{l_1}))
\mathcal{K}^{(n)}(\mathfrak{k}, \mathfrak{l}_1)  
\label{eq:NMarginalIVd} 
\end{eqnarray}

In (\ref{eq:NMarginalI}) we conditioned on the fluctuation 
$\delta \vec{l}_t \in \mathbb{R}^d$
and we integrated over $z \in \mathbb{R}$. In (\ref{eq:NMarginalII})
we assumed that the time $\delta t$ is small enough that at most one jump occurs during it 
and therefore we replaced the conditional fluctuation pdf by the conditional jump pdf
and in (\ref{eq:NMarginalIII}) we used (\ref{eq:ScalingLimitCor}).
In (\ref{eq:NMarginalIV}) we replaced the jump pdf $\varpi$ through
an integral over $\vec{\lambda} \in \mathbb{R}^d$ from the Fourier transform $\tilde{\varpi}$
and we changed the order of integration integrating first over $\vec{\lambda}$ and
then over $\delta \vec{l}_t$.
In (\ref{eq:NMarginalIVa}) 
we defined $\mathfrak{k} := k \sigma^n$, 
$\vec{\mathfrak{l}} := \underline{\underline{\mathfrak{R}}} ((\delta t)^{-\beta \underline{\underline{E}}^{T}} \vec{\lambda})$ 
we introduced an orthogonal operator 
$\underline{\underline{\mathfrak{R}}}$ such that 
$\underline{\underline{\mathfrak{R}}} \vec{\sigma} = (|\vec{\sigma}|,0,\dots,0)$
and we substituted for $\underline{\underline{\mathfrak{R}}} \delta \vec{l}_t$.
In (\ref{eq:NMarginalIVb}) we performed the integrations over 
$(\delta \vec{l}_t)_2,\dots,(\delta \vec{l}_t)_d$
and in the remaining integral over 
$(\delta \vec{l}_t)_1$ we substituted for $w = (\delta \vec{l}_t)_1$.
In (\ref{eq:NMarginalIVc}) we eliminated the delta functions and transformed 
the $d$-dimensional integral over $\vec{\lambda}$ into a one-dimensional integral over 
$\mathfrak{l}_1$.
In (\ref{eq:NMarginalIVd})
we defined a function (kernel) as :
\begin{eqnarray}
\lefteqn{
 \mathcal{K}^{(n)}(\mathfrak{k}, \mathfrak{l}) :=
\frac{1}{2\pi} \int\limits_{-\infty}^\infty dw 
e^{\imath (\mathfrak{k} w^n - \mathfrak{l} w)} =
} 
\label{eq:KernelDefI}\\
&&=
\left\{
\begin{array}{rr}
2 Re\left[ 
e^{\imath \frac{\pi}{2 n}} 
\int_0^\infty d\xi e^{-\mathfrak{k} \xi^n + e^{-\imath \frac{\pi (n-1)}{2 n}} \mathfrak{l} \xi}
\right]
&
\quad\mbox{if $n$ is odd} \\
2 e^{\imath \frac{\pi}{2 n}} 
\int_0^\infty d\xi e^{-\mathfrak{k} \xi^n} \cos\left( \cos(\pi \frac{(n-1)}{2 n}) \mathfrak{l} \xi\right)
&
\quad\mbox{if $n$ is even} 
\end{array}
\right.
\label{eq:KernelDefII}
\end{eqnarray}
and we evaluated it by rotating the range of integration in the complex plane 
by an angle $\pi/(2 n)$ (Wick rotation).
In doing that we applied the Cauchy theorem to a contour composed of 
an interval $w \in [-R, R]$ in the real axis, two arches $w = R e^{\imath \phi}$ for
$\phi \in [0, \pi/(2 n)]$ and  $\phi \in [\pi, \pi + \pi/(2 n)]$ respectively
and a line $w = \xi e^{\imath \pi/(2 n)}$ for $\xi \in [-R,R]$.
The integrals over arches are bounded from above by
$R \int_0^{\pi/(2n)} d\phi \exp\left[ - \mathfrak{k} R^n \sin(n \phi) \pm \mathfrak{l} R \sin(\phi) \right]$ 
for the arch in the upper ($+$) and the lower ($-$) complex half-plane respectively
and they tend to zero for $R\rightarrow \infty$ if $n$ is even and positive.

Now we compute a small $\mathfrak{k}$ expansion of the kernel 
$\mathcal{K}^{(n)}(\mathfrak{k}, \mathfrak{l})$ for $n$ even.  
We denote $a_n := \cos(\pi \frac{(n-1)}{2 n})$, we substitute
$z := \mathfrak{k} \xi^n$ in (\ref{eq:KernelDefII}) and get:
\begin{eqnarray}
\lefteqn{
 \mathcal{K}^{(n)}(\mathfrak{k}, \mathfrak{l}) :=
\frac{e^{\imath \frac{\pi}{2 n}}}{n \mathfrak{k}^{1/n}}
\int_0^\infty dz z^{1/n - 1} \exp\left(-z \right) \cos\left(\frac{a_n l}{\mathfrak{k}^{1/n}} z^{1/n} \right) = }
\label{eq:KernelExpansionI}\\
&&
\frac{e^{\imath \frac{\pi}{2 n}}}{n \mathfrak{k}^{1/n}}
\sum_{m=0}^\infty \frac{(-1)^m}{(2 m)!} \left(\frac{a_n l}{\mathfrak{k}^{1/n}}\right)^{2 m}
\int_0^{\mathfrak{x}} dz z^{\left(\frac{2 m + 1}{n} - 1\right)} \exp(-z) = 
\label{eq:KernelExpansionII}\\
&&
\frac{e^{\imath \frac{\pi}{2 n}}}{n \pi \mathfrak{m}} \left(\frac{ x }{\mathfrak{k}}\right)^{1/n}
\sum_{m=0}^\infty \frac{(-1)^m}{(2 m)!} (\mathfrak{m} \pi)^{2 m + 1}
                  \frac{\Gamma\left( \frac{2 m + 1}{n}, \mathfrak{x} \right)}{ \mathfrak{x}^{\frac{2 m + 1}{n}} } =
\label{eq:KernelExpansionIII} \\
&&
\frac{e^{\imath \frac{\pi}{2 n}}}{\mathfrak{m} \pi}
\exp\left(-\mathfrak{x}\right)
\left( \frac{\mathfrak{x}^{1/n + 1}}{\mathfrak{k}^{1/n}} \right)
\sum_{p=0}^\infty 
\mathfrak{A}_{p+1}
\mathfrak{x}^p
\label{eq:KernelExpansionIV} 
\end{eqnarray}
In (\ref{eq:KernelExpansionII}) we defined 
$\mathfrak{x} := \left( \mathfrak{m} \pi/(2 a_n l) \right)^n \mathfrak{k}$,
we expanded the cosine into a Maclaurin series and,
owing to the fact that the period of the cosine is small for small $\mathfrak{k}$,
the integrand oscillates rapidly and only the vicinity of the singularity of the integrand
contributes meaningful to the integral, we truncated the integration at 
$z = \mathfrak{m} \pi/2$ for some $\mathfrak{m} \in \mathbb{N}$. 
In (\ref{eq:KernelExpansionIII}) we expressed the result 
through the truncated Gamma function.
In (\ref{eq:KernelExpansionIV})  we defined
\begin{equation}
\mathfrak{A}_p := 
\left(
\sum_{m=0}^\infty \frac{(-1)^m}{(2 m + 1)!} 
\frac{\pi^{2 m + 1}}{\prod_{q=1}^p \left(\frac{2 m +1}{n} + q\right)}
\right)
\end{equation}
we made use of a series expansion 
(\ref{eq:GammaFctSeriesExp}) of the truncated Gamma function.
\begin{equation}
\Gamma\left(\alpha,\mathfrak{x}\right) = 
\frac{\mathfrak{x}^\alpha}{\alpha} e^{-\mathfrak{x}}
\sum_{p=0}^\infty \frac{\mathfrak{x}^p}{\prod_{q=1}^p (\alpha + q)}
\label{eq:GammaFctSeriesExp}
\end{equation}
We note that for $\mathfrak{k}$ small enough the kernel satisfies:
\begin{equation}
| \mathcal{K}^{(n)}(\mathfrak{k}, \mathfrak{l}_q) | :=
\frac{\mathfrak{A}_1 e^{-q} q^{1 + 1/n}}{\mathfrak{m \pi}}
\frac{1}{\mathfrak{k}^{1/n}}
\end{equation}
for $\mathfrak{x} = q \Leftrightarrow \mathfrak{l}_q = (\mathfrak{m} \pi)/(2 a_n q^{1/n}) \cdot \left(\mathfrak{k}\right)^{1/n}$
and for $q^{-1} = q(m)^{-1} \in \mathbb{N}$ and $q \le 1$ and thus we have
\begin{equation}
\mbox{lim}_{\mathfrak{k} \rightarrow 0}  
\mathcal{K}^{(n)}(\mathfrak{k}, \mathfrak{l})  \simeq
\delta\left( \mathfrak{l} - \mathfrak{C} \frac{\pi}{2 a_n }\left(\mathfrak{k}\right)^{1/n}  \right)
\label{eq:KernelLimit}
\end{equation}
where $\mathfrak{C} := (\mathfrak{m})/(q(\mathfrak{m})^{1/n})$  is some number.
It is readily seen from (\ref{eq:KernelDefI}) and 
(\ref{eq:KernelDefII}) 
that $\int_{-\infty}^{\infty} d\mathfrak{l}  \mathcal{K}^{(n)}(\mathfrak{k}, \mathfrak{l}) = 1$ and the same holds for the right hand side of (\ref{eq:KernelLimit}).
Inserting (\ref{eq:KernelLimit}) into (\ref{eq:NMarginalIVd})
we obtain (\ref{eq:SmallKValNMargPdf}) {\bf q.e.d.}.

\section{Appendix B}
We calculate the integral (\ref{eq:PolDeltaFct}) for small values of $k$. We have:
\begin{eqnarray}
\lefteqn{
\mathfrak{N}(k, \lambda) := 
\frac{1}{2\pi}
\int_{-\infty}^\infty dz
\exp\left\{\imath k P(e^z - 1)\right\} e^{-\imath \lambda z} =}
\label{eq:PolDeltaFctProofI} \\
&&
\frac{1}{2\pi}
\sum_{\alpha_i} \int_{r_i}^{r_{i+1}}
\frac{d w}{k}
\frac{\left( P^{(-1),i} \left( \frac{w}{k} \right) \right)^{'}}
     {\left[ 1 + P^{(-1),i} \left( \frac{w}{k} \right) \right]}
e^{\imath w}
\exp\left(-\imath \lambda 
           \log\left[ 1 + P^{(-1),i} \left( \frac{w}{k} \right) \right] 
    \right)
\label{eq:PolDeltaFctProofII} \mathop{\rightarrow}_{k \rightarrow 0} \\
&&
\frac{1}{2\pi}
\sum_{\alpha_i} \int_{r_i}^{r_{i+1}}
\frac{d w}{k}
\frac{\left( P^{(-1),i} \left( \frac{w}{k} \right) \right)^{'}}
     {\left[ P^{(-1),i} \left( \frac{w}{k} \right) \right]}
e^{\imath w}
\exp\left(-\imath \lambda \log\left[P^{(-1),i}\left( \frac{w}{k} \right) \right] \right)
\label{eq:PolDeltaFctProofIII} \mathop{\rightarrow}_{k \rightarrow 0} \\
&& =
\frac{1}{2\pi}
\sum_{\alpha_i} \int_{r_i}^{r_{i+1}}
\frac{\imath d w}{k}
\frac{\left( P^{(-1),i} \left( \frac{\imath w}{k} \right) \right)^{'}}
     {\left[ P^{(-1),i} \left( \frac{\imath w}{k} \right) \right]^{1 + \imath \lambda}}
e^{-w}
\label{eq:PolDeltaFctProofV} \\
&& =
\mathfrak{F}_k +
\theta 
\int_0^\epsilon \frac{\imath d w}{k}
\frac{\left( P^{(-1),i} \left( \frac{\imath w}{k} \right) \right)^{'}}
     {\left[ P^{(-1),i} \left( \frac{\imath w}{k} \right) \right]^{1 + \imath \lambda}}
\label{eq:PolDeltaFctProofVI} \\
&& =
\mathfrak{F}_k +
\theta
\int_{ P^{(-1),i}(0) }^{ P^{(-1),i}(\frac{\imath \epsilon}{k}) }
\frac{dz }{z^{1 + \imath \lambda}}
\label{eq:PolDeltaFctProofVII} \\
&& =
\mathfrak{F}_k + \imath \theta
\frac{\left( P^{(-1),i}\left(\frac{\imath \epsilon}{k}\right)\right)^{-\imath \lambda}}{\lambda}
\label{eq:PolDeltaFctProofVIII} \\
&& =
\mathfrak{F}_k + \imath \theta
\frac{\exp( -\imath \lambda \log\left( P^{(-1),i}\left(\frac{\imath \epsilon}{k}\right) \right) )}{\lambda}
\label{eq:PolDeltaFctProofIX}
\end{eqnarray}

In (\ref{eq:PolDeltaFctProofII}) we substituted $w = k P(e^z - 1)$, 
we denoted  by $z_i(y) := P^{(-1),i}(y)$ the $i^{\mbox{th}}$ solution 
(inverse function) 
of the equation
$P(e^z - 1) = y$ and we divided the integration range over $z$ into intervals 
$\left[\alpha_i,\alpha_{i+1}\right]$
in which the function
$P(e^z - 1)$ is monotone and thus invertible.
We also denoted $r_i := k P\left(e^{\alpha_i} - 1\right)$.
In (\ref{eq:PolDeltaFctProofIII}) 
we used the fact that $P(\infty) = \infty$ and thus $P^{(-1)}(\infty) = \infty$
and we approximated the integrand by its large $w/k$ values.

In (\ref{eq:PolDeltaFctProofV}) we rotated the integration line by $\pi/2$ 
in the anti-clockwise direction making use of the fact that the 
integral over a quarter of a circle $w = R e^{\imath \phi}$ for $\phi \in [0,\pi/2]$
is bounded  from the above by $\int_0^{\pi/2} d\phi e^{-R \sin(\phi)}$ 
and it vanishes for $R \rightarrow \infty$.
In (\ref{eq:PolDeltaFctProofVI}) we used the fact that the only root
of $P^{(-1)}(z)$ is at $z=0$ and thus we separated the integral
into two parts the first one equal to $\mathfrak{F} = \mathfrak{F}_k$ 
over the integration range with all singularities
excluded $\mathfrak{U} = \left( \bigcup_i \left[r_i, r_{i+1}\right] \diagdown \left[0,\epsilon\right] \right)$ 
and the second one over $[0, \epsilon]$. 
Note that since $\mbox{lim$_{k \rightarrow 0} |r_{i+1} - r_i| = 0$}$
and the integrand has no singularities in $\mathfrak{U}$
the number $\mathfrak{F}_k \rightarrow 0$ when $k \rightarrow 0$.
The number $\theta$ counts intervals $[r_i,r_{i+1}]$ such that $0 \in [r_i,r_{i+1}]$.  
In (\ref{eq:PolDeltaFctProofVI}) we substituted 
$z = P^{(-1),i}\left(\frac{\imath \epsilon}{k}\right)$ and  
in  (\ref{eq:PolDeltaFctProofVII}) we performed the integral. 
Now we take $\mathfrak{m} > 0$, small $k$ and 
\mbox{$\lambda_{\mathfrak{m}} := 
\left( - \imath \mathfrak{m}^{-1} \right)/
 \left( \log(P^{(-1,i)}\left(\frac{\imath \epsilon}{k}\right) \right)$}
and we evaluate the integral (\ref{eq:PolDeltaFctProofIX}) for this value.
\begin{eqnarray}
\mathfrak{N}(k, \lambda_\mathfrak{m})  := 
\mathfrak{F}_k +
\imath \theta 
\frac{\exp\left( - \mathfrak{m}^{-1} \right)}
     {
\frac{ -\imath \mathfrak{m}^{-1} }
     {\left( \log(P^{(-1,i)}\left(\frac{\imath \epsilon}{k}\right) \right)}
     }
=
\mathfrak{F}_k  
- \mathfrak{m} \exp\left( - \mathfrak{m}^{-1} \right)
\theta 
\left( \log(P^{(-1,i)}\left(\frac{\imath \epsilon}{k}\right) \right)
\end{eqnarray}
We fix $k$ and let $\mathfrak{m} \rightarrow 0$.
Then $\lambda_{\mathfrak{m}} \rightarrow 0$ and we see that
$\mathfrak{N}(k, 0)  \rightarrow 0$.
Furthermore from 
(\ref{eq:PolDeltaFctProofI}) we have 
$\int_{-\infty}^\infty d\lambda \mathfrak{N}(k, \lambda) = 1$
we conclude that 
\begin{equation}
\mbox{lim}_{k \rightarrow 0} \mathfrak{N}(k, \lambda) = 
\delta\left(\lambda + 
\frac{\imath \mathfrak{m}^{-1} }
{\left( \log(P^{(-1,i)}\left(\frac{\imath \epsilon}{k}\right) \right)}\right)
\label{eq:LimLogDeltaAppend}
\end{equation}
and the number $\mathfrak{m}$ can be evaluated from
\begin{equation}
\left|\lambda_\mathfrak{m}\right| \mathfrak{N}(k, \lambda_\mathfrak{m}) 
= e^{-m^{-1}}\theta = \frac{1}{2}
\label{eq:DeltaConstEval}
\end{equation}
{\bf q.e.d.}.

\end{document}